\documentclass[12pt]{article} 
\usepackage{graphicx,floatflt,amssymb,rotate}

\def\gtap{\raisebox{-.4ex}{\rlap{$\sim$}} \raisebox{.4ex}{$>$}}

\begin{document}

\vskip 30pt  
 
\begin{center}  
{\Large \bf LHC limits on KK-parity non-conservation in the
strong sector of  universal extra-dimension models} \\
\vspace*{1cm}  
\renewcommand{\thefootnote}{\fnsymbol{footnote}}  
{ {\sf Anindya Datta\footnote{email: adphys@caluniv.ac.in}}, 
{\sf Amitava Raychaudhuri\footnote{email: palitprof@gmail.com}},  
{\sf Avirup Shaw\footnote{email: avirup.cu@gmail.com}} 
} \\  
\vspace{10pt}  
{\small {\em Department of Physics, University of Calcutta,  
92 Acharya Prafulla Chandra Road,\\ Kolkata 700009, India}}\\ 
   
\normalsize

\end{center}

\begin{abstract}  
\noindent

In five-dimensional universal extra-dimensional models
compactified on an $S^1/Z_2$ orbifold four-dimensional kinetic
terms  are allowed at the two fixed points. If these terms are
unequal then Kaluza-Klein (KK) parity is broken.  Within such a
framework we consider resonant production of the $n = 1$
KK-gluon at the  LHC and its subsequent decay to
$t\bar{t}$, where both production and decay are KK-parity
non-conserving.  We use, for the first time, the exclusion
data for a $t\bar{t}$ resonance obtained by the  LHC experiments
to limit the  mass range of the lowest gluon
excitation and, in a correlated fashion, of the $n = 1$ quark
excitation of the KK-parity-violating model which are both found to
be in the ballpark of  600 - 2000 GeV. 
\\
 
\vskip 5pt \noindent  
\texttt{PACS Nos:~11.10.Kk, 14.80.Rt, 13.85.-t  } \\  
\texttt{Key Words:~~Universal Extra Dimension, Kaluza-Klein, LHC}  
\end{abstract}

\renewcommand{\thesection}{\Roman{section}}  
\setcounter{footnote}{0}  
\renewcommand{\thefootnote}{\arabic{footnote}}  

\section{Introduction}

The detection of a Higgs-like scalar particle at the Large Hadron
Collider (LHC) at CERN is a landmark accomplishment. Much
activity is now aimed to uncover detailed properties of this new
state and compare it with the expectations from the standard model
(SM).  There is also a  continuing  interest regarding the
physics which lies beyond the standard model. The evidence for
such physics, albeit indirect, can be traced to the issue of
naturalness of the Higgs-scalar mass, the observed masses and mixing
of neutrinos, and the quest for a dark matter candidate. The
energy scale for new physics remains unknown but there are
several motivations which encourage us to expect that it may
well be within the reach of the LHC.  Here our intention is to
constrain a class of non-minimal universal extra-dimensional
models where the lowest ($n = 1$) Kaluza-Klein excitations are
not stabilized by any symmetry.  We show that the  data
reported by the ATLAS \cite{atlas7T, atlas8T} and the CMS
\cite{cms7T, cms8T} Collaborations  
excluding a heavy resonance in the {$t\bar{t}$}
channel eliminate significant regions in the $n = 1$
KK-gluon and KK-quark mass plane.

In the simplest Universal Extra Dimension (UED) model there is
one extra flat spacelike dimension and it is accessible to all
particles \cite{acd}.  The extra dimension $y$ is compact (radius
of compatification $R$) and has a $Z_2$ symmetry ($y \rightarrow
-y$) to incorporate chiral fermions. For every SM particle one
has a tower of KK excitations, each member being specified by an
integer $n = 0,1 ,2, \ldots$, the standard model particle being
the $n = 0$ mode of the tower. The SM masses of the particles are
small compared to $1/R$ and it is a good approximation to take
the  KK states for all particles at any level $n$ to be
degenerate with a mass $n/R$.

The $Z_2$ symmetry produces a conserved KK-parity which is
$(-1)^n$ for the $n$-th KK-level. The SM particles are of even
parity while those of the first level are odd. KK-parity ensures
that the lightest among the $n = 1$ particles is absolutely
stable and so can be a  dark matter candidate, the Lightest
Kaluza-Klein Particle (LKP). This is the essence of the Universal
Extra Dimension (UED) Model.

The above $S^1/Z_2$ orbifold compactification has two fixed
points at $y = 0$ and $y = \pi R$.  At these boundary points 
inclusion of additional four-dimensional interactions is allowed by
the symmetry. In fact,
these terms are useful as counterterms for compensating
loop-induced contributions \cite{georgi} of the five-dimensional
theory. In the simplest choice, the minimal Universal
Extra-Dimensional Models (mUED) \cite{cms1, cms2}, these terms
are chosen so that they exactly cancel the five-dimensional loop
effects at the cutoff scale of the theory $\Lambda$. Thus these boundary
contributions, e.g., logarithmic corrections to
masses of KK particles, are such that they vanish at the scale
$\Lambda$. At lower energies these  contributions remove the mass degeneracy
among states at the same KK-level $n$.

The radius of compactification $R$ sets the mass scale of the
theory and splittings within the KK-states of the same level are
controlled by the cutoff $\Lambda$. They can be constrained by
using known measurement results. Thus, from the muon $(g-2)$ \cite{nath},
flavour changing neutral currents \cite{chk,buras,desh}, $Z \to
b\bar{b}$ decay \cite{santa}, the $\rho$ parameter
\cite{acd,appel-yee}, and other electroweak precision tests
\cite{ewued, precision}, one has typically $1/R~\gtap~300-600$ GeV.  
Comparing the production and leptonic decay of $n = 2$
electroweak gauge bosons with the CMS LHC data a limit of
$1/R~\gtap~715$ GeV has been placed \cite{flacke2}.

In this work we explore the scenario where  the boundary terms
depart from their special choice of mUED. In addition, the
boundary terms at the two fixed points are allowed to be unequal.
This leads to non-conservation of KK-parity and opens the door
for $n = 1$ KK-states to be produced singly in SM particle
collisions and also for them to decay to $n = 0$
states\footnote{This is similar to R-parity violating
interactions in supersymmetry.} some of which has been emphasized
in an earlier work \cite{ddrs}. Because of these
features the picture considered  here is termed KK-parity
violating UED.  

Here we examine the coupling of the $n = 1$ KK-gluon to a pair of
SM-quarks ($n =0$ states).  Such couplings are absent in mUED and
are hallmarks of the non-minimality discussed above. They provide
an avenue for comparing the predictions of the theory with
measurements at the LHC.

Specifically, we consider resonant production of the $n = 1$
KK-gluon in $pp$ collisions at the LHC through the KK-parity
non-conserving coupling. We take the KK-gluon to be the lightest
$n = 1$ level particle. Once produced, KK-conserving decays being
kinematically disallowed, the KK-gluon  decays to a pair of
zero-mode quarks through the same KK-parity-violating coupling.
The branching ratio is equal for all types of quarks and hence it
is 1/6 for the $t \bar{t}$ mode which we examine.  Both the
ATLAS and CMS collaborations have looked for the signal of a
resonance produced in $pp$  collisions which decays to a pair of
top-antitop quarks. Results from ATLAS for the 7 TeV
\cite{atlas7T} and 8 TeV \cite{atlas8T} LHC runs are now available 
as are the corresponding findings of CMS in \cite{cms7T} and
\cite{cms8T} respectively.   From the lack of observation of
such a state, 95\% C.L. upper limits on the cross section times
branching ratio of such a signal as a function of the resonance
mass have been reported in these publications.  Here we use the limits
from the 8 TeV runs to constrain the masses of the $n = 1$
level KK quarks and gluons of the model. We would like to
emphasize that this is the first effort to restrict the
parameters of KK-parity-violating UED using existing LHC data.

The two essential ingredients for calculating the signal are the
mass of the $n = 1$ gluon state and the strength of its
KK-parity-violating couplings. In the following section we
briefly review the UED scenario with boundary-localized kinetic
terms and lead up to the KK-excitation masses in such a
framework.  In the next section we calculate the
$Z_2$-parity-violating coupling involving the first excitation of
the KK-gluon and a zero-mode quark-antiquark pair. With these
results we then derive the expected $t\bar{t}$ signal from the
production of the KK-gluon at the LHC and its subsequent decay.
This is compared with the  CMS \cite{cms8T} and ATLAS
\cite{atlas8T} 8 TeV results  and the
restrictions on the KK-excitation masses are exhibited.
Our conclusions appear at the end.

\section{KK-parity-violating UED, KK masses}

In nonminimal UED one can consider kinetic and mass terms localized at the
fixed points. Here we restrict ourselves to boundary-localized
kinetic terms only \cite{Dvali} - 
\cite{asesh}, \cite{ddrs}.  Thus we consider a
five-dimensional theory with additional four-dimensional kinetic
terms at the boundaries at $y=0$ and $y=\pi R$.

We illustrate the idea by considering free fermion fields
$\Psi_{L,R}$ whose zero modes are the chiral projections of the
SM fermions. The five-dimensional action with BLKT is \cite{schwinn}
\begin{eqnarray} 
S & = \int d^4x ~dy \left[ \bar{\Psi}_L i \Gamma^M \partial_M \Psi_L 
+ r^a_f \delta(y) {\phi} ^\dagger _L i \bar \sigma^\mu \partial_\mu \phi_L 
+ r^b_f \delta(y - \pi R) {\phi} ^\dagger _L i \bar \sigma^\mu
\partial_\mu \phi_L
\right. \nonumber \\
&  \left. + \bar {\Psi} _R i \Gamma^M \partial_M \Psi_R
+ r^a_f \delta(y) {\chi} ^\dagger _R i {\sigma}^\mu \partial_\mu \chi_R 
+ r^b_f \delta(y - \pi R) {\chi} ^\dagger _R i {\sigma}^\mu
\partial_\mu \chi_R
\right]  .
\label{faction}
\end{eqnarray} 
Here $\sigma^\mu \equiv (I, \vec{\sigma})$ and $\bar{\sigma}^\mu
\equiv (I, -\vec{\sigma})$, $\vec{\sigma}$ being the $(2 \times
2)$ Pauli matrices.  $r^a_f, r^b_f$ are the strengths
of the boundary terms which are chosen to be the same for $\Psi_L$
and $\Psi_R$ for simplicity.

It is helpful to express five-dimensional fermion fields
using two component chiral spinors\footnote{The Dirac gamma
matrices are in the chiral representation with $\gamma_5 = 
diag(-I, I)$.} \cite{schwinn}:
\begin{equation} 
\Psi_L(x,y) = \pmatrix{\phi_L(x,y) \cr \chi_L(x,y)} 
=   \sum^{\infty}_{n=0} \pmatrix{\phi_n(x) f_L^{n}(y) \cr \chi_n(x) g_L^{n}(y)}
\;\; , 
\label{fiveDL}
\end{equation} 
\begin{equation} 
\Psi_R(x,y) = \pmatrix{\phi_R(x,y) \cr \chi_R(x,y)} 
=   \sum^{\infty}_{n=0}  \pmatrix{\phi_n(x) f_R^{n}(y) \cr \chi_n(x) g_R^{n}(y)} 
\;\;  . 
\label{fiveDR}
\end{equation} 
Below we examine the case of $\Psi_L$ in detail. The results 
for $\Psi_R$ will be similar and can be obtained by
making appropriate changes.

Variation of the action functional Eq.  (\ref{faction}) utilising
Eq. (\ref{fiveDL})  results in coupled equations for the
$y$-dependent wave-functions, $f_L^{n}, g_L^{n}$. Following
routine steps\footnote{More details in the same notations and
conventions can be found in \cite{ddrs}.} one finds:
\begin{equation}
\left[1 + r^a_f \delta(y) + r^b_f \delta(y - \pi R) \right] m_n f_L^n - 
\partial_y g_L^n = 0,\;\;
m_n g_L^n + \partial_y f_L^n = 0, \; (n = 0,1,2, \ldots).
\end{equation}
Eliminating $g_L^n$ one obtains the equations:
\begin{eqnarray}
\partial_y^2 f_L^n &+& \left[1 + r^a_f \delta(y) + r^b_f \delta(y - \pi R) 
\right] m_n^2 f_L^n = 0 .
\end{eqnarray}
From now onwards we drop the subscript $L$ on the wave-functions  and
denote them simply by $f$ and $g$.

The boundary conditions are \cite{carena}
\begin{equation}
f^n(y)|_{0^-} = f^n(y)|_{0^+},\;\; f^n(y)|_{\pi R^+} = f^n(y)|_{\pi R^-} , 
\end{equation}
\begin{equation}
\frac{df^n}{dy}\bigg|_{0^+} - \frac{df^n}{dy}\bigg|_{0^-} = -r_f^a
m_n^2 f^n(y)|_{0}, \;\;
\frac{df^n}{dy}\bigg|_{\pi R^+} - \frac{df^n}{dy}\bigg|_{\pi R^-} = -r_f^b
m_n^2 f^n(y)|_{\pi R} .
\end{equation}
One then obtains the solutions:
\begin{eqnarray}
f^n(y) &=& N_n \left[ \cos (m_n y) - \frac{r_f^a m_n}{2} \sin (m_n
y) \right] \;,\;\;  0 \leq y < \pi R,   \nonumber \\ 
f^n(y) &=& N_n \left[ \cos (m_n y) + \frac{r_f^a m_n}{2} \sin (m_n
y) \right] \;,\;\; -\pi R \leq y < 0.
\label{sol1}
\end{eqnarray}
where the masses $m_n$ for  $n = 0,1, \ldots$ 
are obtained from the transcendental equation \cite{carena}:
\begin{equation} 
(r^a_f r^b_f ~m_n^2 - 4) \tan(m_n \pi R)= 2(r^a_f + r^b_f) m_n \;.
\label{trans1}
\end{equation}

The solutions are {\em orthonormal}, i.e.:
\begin{equation}
\int dy \left[1 + r^a_f \delta(y) + r^b_f \delta(y - \pi R)
\right] ~f^n(y) ~f^m(y) = \delta^{n m},\;\;
\end{equation}

These wave-functions are combinations of a sine
and a cosine function unlike in the case of mUED where they are 
one or the other of these two trigonometric functions. This
difference and that the KK masses are solutions of Eq. (\ref{trans1})
rather than just $n/R$ are at the root of the distinguishing features of
this model.

In this paper we examine two versions of KK-parity-violating UED.
In one we take symmetric boundary-localized terms for fermions,
i.e., $r^a_f = r^b_f \equiv r_f$. The other case has the BLKT at
one of the fixed points only:   $r^a_f \neq 0, ~r^b_f = 0$. In
this second alternative Eq. (\ref{trans1}) becomes
\begin{equation} 
\tan(m_n \pi R)=-\frac{r^a_f m_n}{2} .
\label{trans3f}
\end{equation}

The normalisation constant $N_n$ is determined from
orthonormality. When $r^a_f = r^b_f \equiv r_f$  
\begin{equation}
 N_n = \sqrt{\frac{2}{\pi R}}\left[ \frac{1}{\sqrt{1 + \frac{r_f^2 m_n^2}{4} 
+ \frac{r_f}{\pi R}}}\right].
\label{norm1}
\end{equation}
For the other case when $r^b_f = 0$  we use $r^a_f \equiv r_f$
and one has
\begin{equation}
 N_n = \sqrt{\frac{2}{\pi R}}\left[ \frac{1}{\sqrt{1 + \frac{r_f^2 m_n^2}{4} 
+ \frac{r_f}{2 \pi R}}}\right].
\label{norm2}
\end{equation}
In this work we deal only with the zero modes and the $n=1$
KK wave-functions of the five-dimensional fermion fields.

The action for the five-dimensional gluon field, $G_N ~(N = 0
\ldots 4)$ with BLKT $r^a_g, r^b_g$ at the fixed points can be
similarly written down. It is straightforward to show following
similar steps\footnote{These steps are discussed in detail in
\cite{ddrs}.} that in the $G_4 = 0$ gauge the gluon field has the
KK-expansion:  
\begin{equation} 
G_{\mu}(x,y)=\sum^{\infty}_{n=0}G_{\mu}^{n}(x) a^n(y),
\end{equation} 
where the functions $a^n(y)$  are of the same form as Eq.
(\ref{sol1}). In this case the five-dimensional contributions to
the KK-gluon mass, $m_n$, satisfy
\begin{equation}
(r_g^a r_g^b m_{n}^{2}-4)~\tan \left(m_{n}\pi R\right) = 2(r_g^a+r_g^b)
m_{n} \; .
\label{trans2}
\end{equation}
which is identical to Eq. (\ref{trans1}) for fermions.

For the other case that we also consider ($r^a_g \neq 0 , \;
r^b_g = 0$) the transcendental equation (\ref{trans2}) reduces to 
\begin{equation} 
\tan(m_n \pi R)=-\frac{r^a_g m_n}{2} .
\label{trans3}
\end{equation} 
This equation is the same as Eq. (\ref{trans3f}) for fermions
with similar BLKT.

As the KK-masses of fermions and gauge bosons are obtained from
similar equations it is convenient to discuss the solutions
together. Below we use $r_\alpha^a, r_\alpha^b$ to stand for the
BLKT strengths with $\alpha = f$ or $g$. Our focus will be on the
$n = 1$ state.

\begin{figure}[h] 
\begin{center} 
{\hskip -9cm}
\includegraphics[scale=0.50, angle=270]{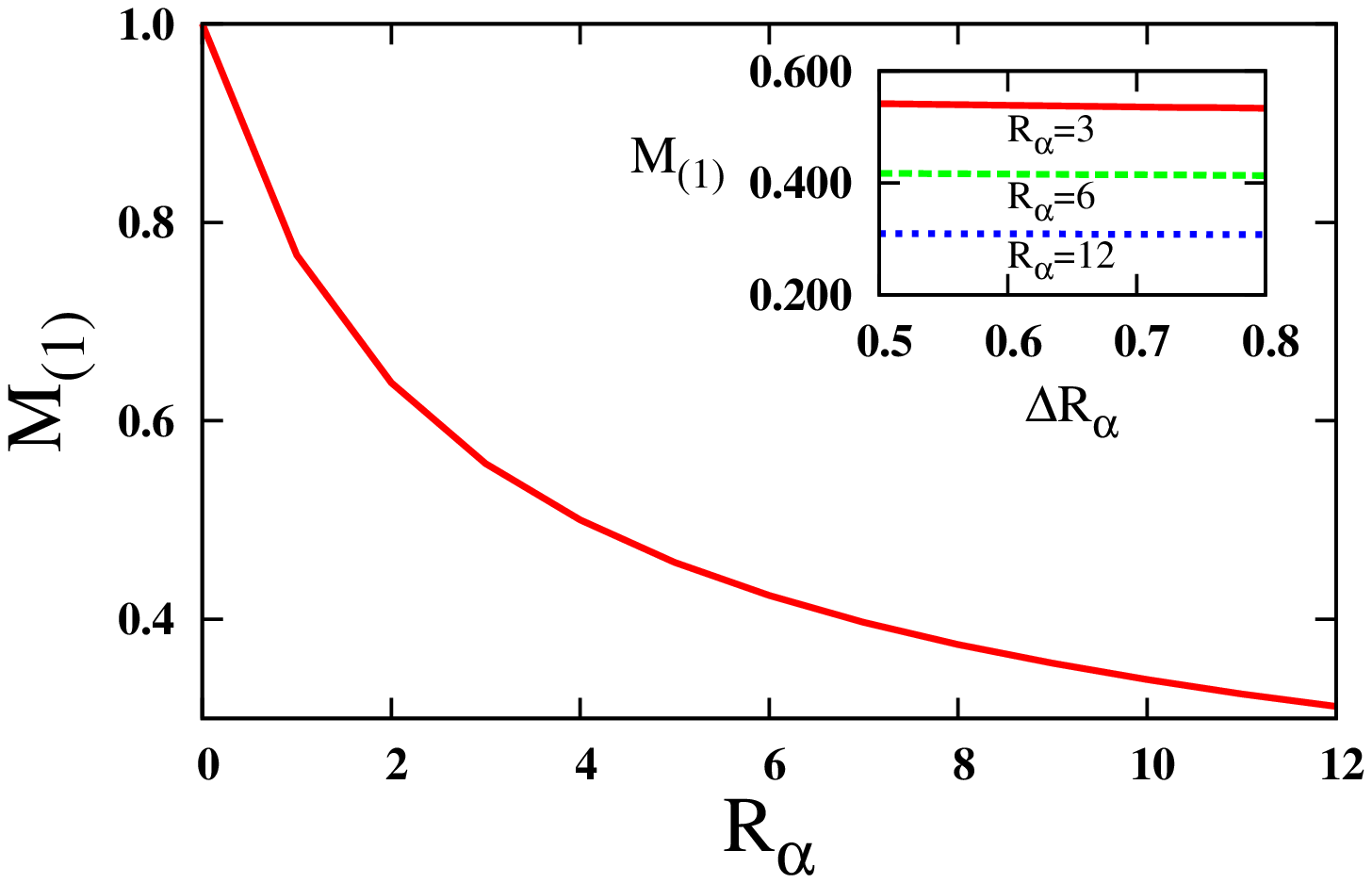}
{\vskip -5.0cm}
{\hskip 7cm}
\includegraphics[scale=.50, angle=270]{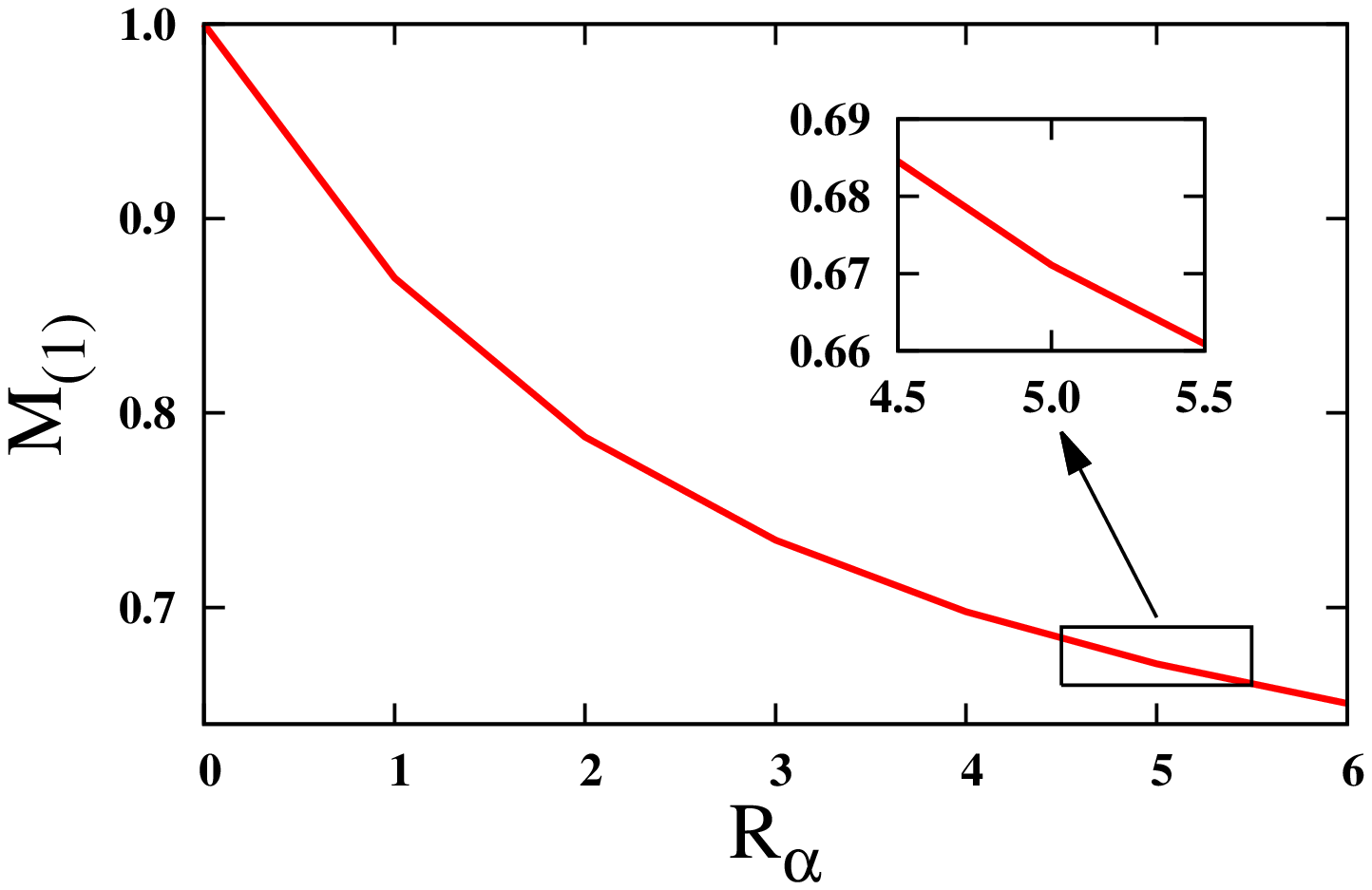}
\caption{Left panel: $M_{(1)}\equiv m_{\alpha^{(1)}} R$ as a
function of $R_\alpha \equiv r_\alpha^a/R$ when $r_\alpha^a =
r_\alpha^b$. In the inset is shown the dependence of $M_{(1)}$ on
${\Delta R_\alpha} \equiv (r_\alpha^b - r_\alpha^a)/R$ for
several $R_\alpha$. Right panel: $M_{(1)}$ as a function of
$R_\alpha$ when the BLKT is present only at the $y = 0$ fixed
point. The inset shows a blow-up of the region of $R_\alpha$ that is
considered later. Note the very mild variation of $M_{(1)}$ in
the insets of both panels.  The plots apply for $\alpha = f$
(fermions) and $g$ (gluons).}
\label{KKmass} 
\end{center} 
\end{figure}


In Fig. \ref{KKmass} we show the variation of $M_{(1)}
\equiv m_{\alpha^{(1)}} R$, which is dimensionless, in the two alternate
cases. In the left panel $M_{(1)}$ is shown as a function of
$R_\alpha \equiv r_\alpha^a/R$ in the symmetric ($r_\alpha^a =
r_\alpha^b$) limit. In the inset is presented the dependence of
$M_{(1)}$ on the asymmetry parameter $\Delta R_\alpha$, for
several choices of $R_\alpha$.  The range of variation of $\Delta
R_\alpha$ shown in the inset is what we use in our later
discussion.  It is seen that   $M_{(1)}$  changes very slowly.
Note also that the mass of the $n = 1$ state for a particular
$R_\alpha$ always remains more than that corresponding to any
larger $R_\alpha$ for the entire variation of $\Delta R_\alpha$.
Therefore, irrespective of the value of $\Delta R_\alpha$, the
mass ordering within the $n = 1$ level is determined on the basis
of $R_\alpha$.  In the right panel, we present  $M_{(1)}$ as a
function of $R_\alpha=r^a_\alpha/R$  when the BLKT is present
only at $y = 0$. In the inset is shown the region in which we
later choose the gauge boson BLKT. Here too the variation of
$M_{(1)}$ is hardly significant.  The key message from both
panels is that the KK-mass falls with increasing $R_\alpha$; the
fermion or gauge boson with the largest $R_\alpha$ is the
lightest $n = 1$ KK state.  In this work we keep $R_g > R_f$
to ensure that the $n = 1$ gluon is lighter than the quarks of
the same level. So, the former cannot decay {\em via} KK-number
conserving modes (which would have dominated if allowed) and the
branching ratio to $t\bar{t}$ is 1/6.

\section{Coupling of $G^{1}$ with zero-mode quarks} 

Besides the masses, the other ingredient required for the proposed 
calculation is the strength of the $G^1 f^0 f^0$ coupling. It is
given by
\begin{eqnarray} 
g_{G^{1}f^{0}f^{0}} &=&  
g_5(G)  ~\int^{\pi R}_{0} 
(1+r_{f}\{\delta(y)+\delta(y-\pi R)\})
f_{L}^{0}f_{L}^{0}a^{1} dy  \nonumber  \\
&=& g_5(G) ~\int^{\pi R}_{0} (1+r_{f}\{\delta(y)+\delta(y-\pi R)\})
g_{R}^{0}g_{R}^{0}a^{1} dy  
\label{coup0}
\end{eqnarray}
The five-dimensional gauge coupling $g_5$ which appears above
is related to the usual coupling $g$ through 
\begin{equation}
g_5 = g ~\sqrt{\pi R ~S_G}  
\end{equation}
with
\begin{equation}
S_{G} = \left(1+\frac{R^a_g+ R^b_g}{2\pi}\right). 
\end{equation}
As noted in the previous section, the wave-functions $f_L^0,
g_R^0$ for zero-mode quarks  and $a^1$ for the KK-gluon depend
on the choices made for the boundary localized terms.

We remark in passing that, irrespective of the nature of the
gluon boundary terms, the coupling $g_{G^{1}G^{0}G^{0}}$ is always
zero. Thus the resonant production of $G^1$ is initiated only by
quarks and antiquarks in the colliding particles and the gluonic
content of the proton plays no role.

\subsection{BLKT at both fixed points}

The first option which we study has BLKTs of the same  strength
at the two fixed points for quarks ($r_f^a = r_f^b = r_f$) 
while for the gauge bosons ($r_g^a \neq r_g^b$).
The $y$-dependent wave-functions of our interest here are found to be
\begin{equation}
f_{L}^{0} = g_{R}^{0} = \frac{1}{\sqrt{\pi R(1 + R_f/\pi)}}, 
\end{equation}
and
\begin{equation}
a^{1} = N_{G}^{1} 
\left[\cos \left( \frac{M_{(1)}y}{R} \right)-\frac{R^{a}_g M_{(1)}}{2}
\sin \left(\frac{M_{(1)}y}{R}\right)\right],
\end{equation}
with
\begin{equation}
N_{G}^{1} = \sqrt{\frac{1}{\pi R}}~\sqrt{\frac{8(4+M_{(1)}^{2}{R^b_g}^2)}
{2\left(\frac{R^{a}_g+R^{b}_g}{\pi}\right)(4+M_{(1)}^{2}R^{a}_gR^{b}_g)
+(4+M_{(1)}^{2}{R^{a}_g}^2)(4+M_{(1)}^{2} {R^{b}_g}^2)}}~~, 
\end{equation}
where  we have used as earlier $M_{(1)} \equiv m_{g^{(1)}} R$, and
the scaled dimensionless variables
\begin{equation}
R_f \equiv r_f/R, \;\;  R^a_g \equiv r^a_g/R, \;\; {\rm and} \;\;
R^b_g \equiv r^b_g/R.
\end{equation}
Using the above  we get
\begin{eqnarray} 
g_{G^{1}f^{0}f^{0}} &=&  
 \frac{g(G) \sqrt{\pi RS_G} }{\left(1+\frac{R_{f}}{\pi}\right)}
\;N^1_G\;\left[\frac{\sin(\pi M_{(1)})}
{\pi M_{(1)}}\left\{1-\frac{M_{(1)}^{2}R^{a}_gR_{f}}{4}\right\}\right.
\nonumber \\
&&
\left. +\frac{R^{a}_g}{2\pi}\left\{\cos(\pi M_{(1)})-1\right\}+
\frac{R_{f}}{2\pi}\left\{\cos( \pi M_{(1)})+1\right\}\right] .
\label{coup1}
\end{eqnarray}

In the left panel of Fig. \ref{KKcoupling} we plot the square
of the coupling for a fixed value\footnote{We have checked that
the results are not dramatically different for the other values
of $R^a_g$ that we consider later.} of $R^a_g= 3.0$ 
as a function of $R_f$ for several values of $\Delta
R_g$. It is seen that the strength of the coupling decreases
as $R_f$ increases while it increases as $\Delta R_g$ increases.

Physics consequences of these couplings are discussed in the
next section. At this stage we urge the reader to note that the
KK-parity-violating coupling gets smaller as $R_f$ tends
towards $R_g$ i.e.,  as the splitting among the $n = 1$
KK-excitations is decreased.  Also,  it can be readily seen using
Eq. (\ref{trans2}) that if $R^a_g = R^b_g$, i.e., the BLKTs are
symmetric at $y = 0$ and $y = \pi R$ for  the gauge boson, as
chosen for the quarks, the coupling in Eq. (\ref{coup1})
vanishes. This can be traced to  a $y \longleftrightarrow (y-\pi
R)$ $Z_2$-symmetry of the theory for this choice which forbids an
$n = 1$ state to couple exclusively to zero modes. In general, 
$g_{G^{1}f^{0}f^{0}}$ decreases as $\Delta R_g$ gets smaller.

\subsection{BLKT at one fixed point}

In the second case, for both the quarks and the gauge bosons we
assume that the BLKT are present at only the $y=0$ fixed point. The
$y$-dependent wave-functions in this case are
\begin{equation} 
f_L^{0} = g_R^{0} = \frac{1}{\sqrt{\pi R(1 + R_f/2 \pi)}}, 
\end{equation}
and
\begin{equation}
a^{1} = \sqrt{\frac{1}{\pi R}}~\sqrt\frac{2}{1+\left(\frac{R_g
M_{(1)}}{2}\right)^2+\frac{R_g}{2\pi}}
\left[\cos\left(\frac{M_{(1)}y}{R}\right)-
\frac{R_g M_{(1)}}{2}\sin\left(\frac{M_{(1)} y}{R}\right) 
\right] \;,
\end{equation} 
where $R_g \equiv r_g/R$. With these wave-functions we get for this case
\begin{eqnarray} 
g_{G^{1}f^{0}f^{0}} &=&  
\frac{\sqrt{2} ~g(G) \sqrt{S_G}}
{\left(1+\frac{R_{f}}{2\pi}\right) 
\sqrt{1+\left(\frac{R_g M_{(1)}}{2}\right)^2+\frac{R_g}{2\pi}}}  
\left(\frac{R_{f}-R_g}{2\pi}\right) \;.
\label{coup2}
\end{eqnarray} 

\begin{figure}
\begin{center}
{\hskip -9cm}
\includegraphics[scale=.50, angle=270]{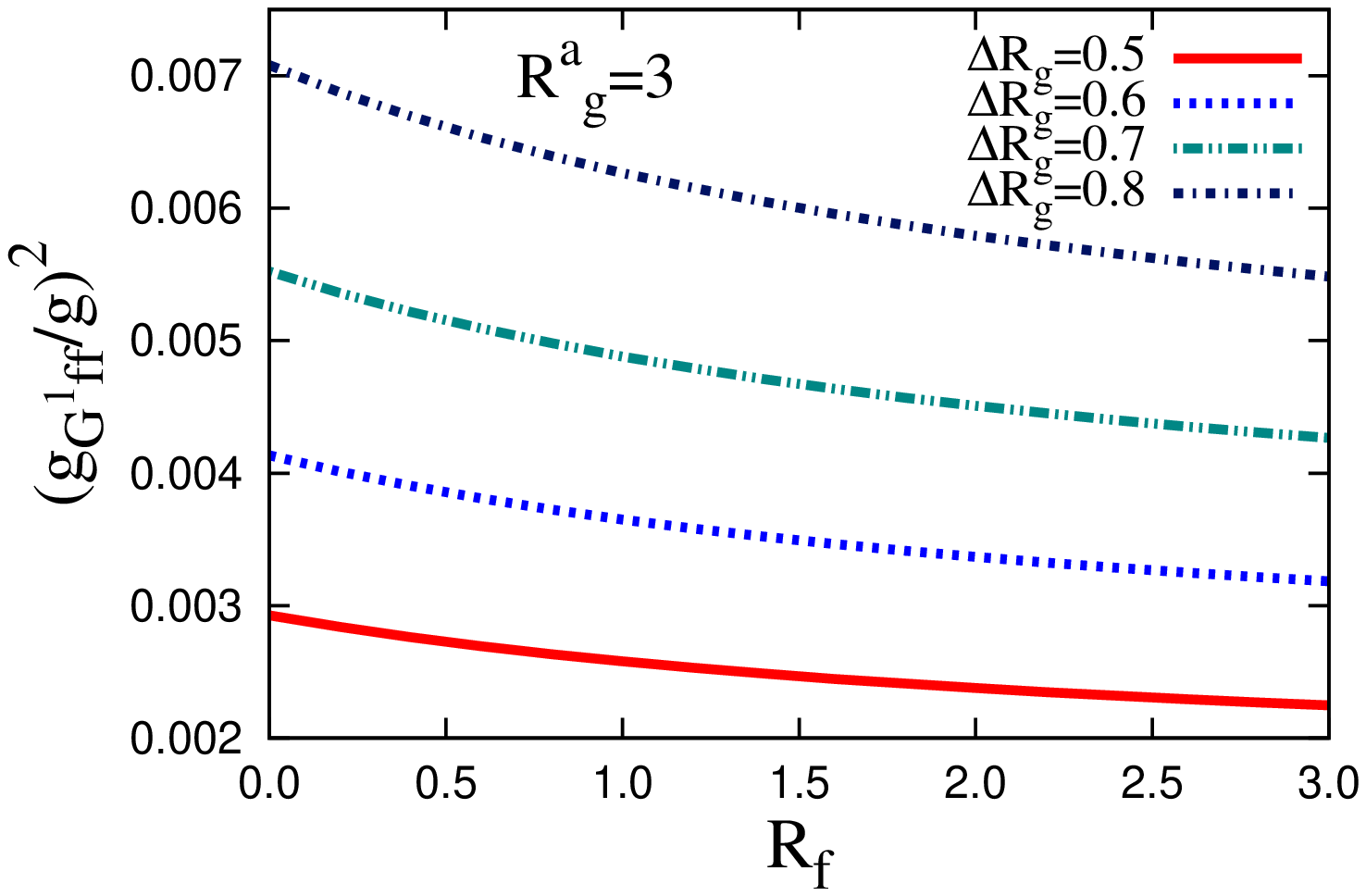}{\vskip -5.4cm}
{\hskip 6cm}
\includegraphics[scale=1.0, angle=0]{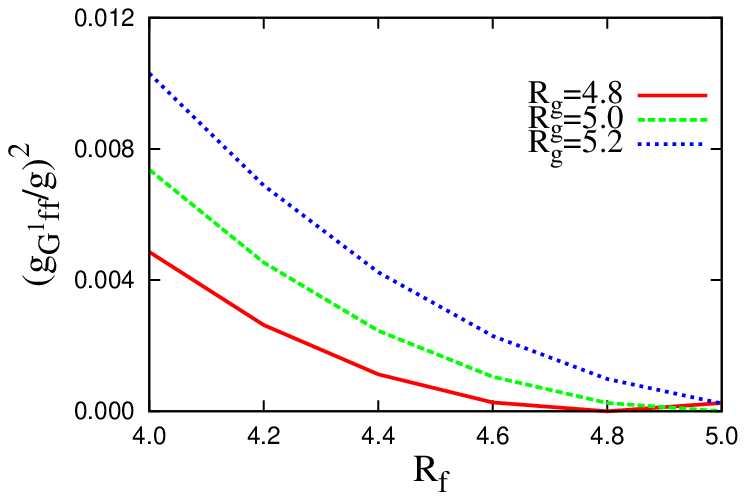}
\caption{Left panel: Variation of the scaled KK-parity-violating
coupling squared between $G^1$ and a pair of zero-mode quarks with
$R_f \equiv R_f^a = R_f^b$ for several $\Delta R_g$, for  $R^a_g=3.0$.
Right panel: Variation of the same coupling with
$R_f$ for different choices of $R_g$ when the quark and boson
BLKTs are present only at the $y = 0$ fixed point.}
\label{KKcoupling} 
\end{center} 
\end{figure}

In the right panel  of Fig. \ref{KKcoupling} we plot the
square of the coupling strength as a function of $R_f$ for
several choices of $R_g$ for this alternative. In order to keep
the $n = 1$ KK-gluon lighter than the quarks of the same level
we have kept $R_g > R_f$. It is to be noted that the strength of
the coupling decreases as $R_f$ increases but it increases as
$R_g$ increases. The coupling vanishes if $R_g=R_f$, as can be seen from Eq.
(\ref{coup2}).

\section{Single $G^{1}$ production and decay to $t\bar{t}$} 

From here onwards we will not explicitly write the KK-number $(n
= 0)$ as a superscript for the SM particles.  Wave-functions with
no superscripts are for the SM particles.

We now investigate the single production of $G^1$ at the LHC.  We
consider the process $p p ~(q \bar q)
\rightarrow G^{1}$ followed by $G^{1} \rightarrow t\bar{t}$. Note
that both the production and decay of $G^1$ involves the
KK-parity-violating coupling. The signature of this mode would be
a resonance in the $t\bar{t}$ channel at the $G^1$ mass.  The
ATLAS and CMS collaboration have both searched for such a
resonance in $pp$ collisions at 7 and 8 TeV \cite{atlas7T,
atlas8T, cms7T, cms8T}.  From non-observation of such a signal
95\% C.L. upper bounds have been placed on the cross section
times  branching ratio as a function of the mass of a $t
\bar{t}$ resonance. Comparing these bounds with the calculated
values in the KK-pariy-violating  framework enables the
restriction of the parameter space of the model. To get the most
up-to-date bounds we use the latest 8 TeV results. At this energy
CMS has published \cite{cms8T} the analysis of 19.7 $fb^{-1}$ of
data while ATLAS has presented \cite{atlas8T} bounds from 14.3
$fb^{-1}$ of data. We use the former in our considerations below
but also remark on the constraint following from the latter.

The key quantities here are the KK-gluon mass and its coupling to
zero-mode quarks. In non-minimal UED, the mass of $G^1$ is
determined from Eqs.  (\ref{trans2}) and (\ref{trans3}) by $1/R$
and the gluon BLKT  $R^{a,b}_{g}$.  The resonance masses
excluded by the ATLAS and CMS results are bounds on the $n = 1$
gluon mass in this model. This restricts $1/R$ and $R^{a,b}_g$.
Further, the single production and the decay of $n=1$
KK-excitations of gluons to SM quarks are driven by
KK-parity-violating couplings which depend on the quark BLKT
$R_f$ and also vanish unless the strengths of the gauge BLKT
parameters localized at the two fixed points are different, i.e.,
$\Delta R_g \neq 0$.  A specific upper bound on the event
rate as quoted by  CMS \cite{cms8T} therefore translates to
constraints on the above parameters and thence to the mass of the
KK-excitations of quarks.

As noted earlier, in spite of the onset of KK-parity violation
the  coupling $g_{G^{1}G G}$ vanishes identically. Consequently,
the production of $G^1$ in $pp$ collisions is driven solely by $q
\bar q$ fusion.  An analytic expression for the resonant
production cross section of the $n = 1$ KK-gluon from $q \bar q$
fusion in the collision of two protons can be expressed in a
compact form:
\begin{equation}
\sigma (p p \rightarrow G^{1} + X) = \frac{4 \pi^2}{3 m_{g^{(1)}}^3}\;\sum_i 
\Gamma(G^{1} \rightarrow q_i \bar q_i)\;\int_\tau ^1 \frac{dx}{x}\;
\left[f_{\frac{q_{i}}{p}}(x,m_{g^{(1)}}^2) 
f_{\frac{{\bar q_{i}}}{p}}(\tau/x,m_{g^{(1)}}^2) + 
q_i \leftrightarrow \bar q_i \right]
\label{x-sections}
\end{equation}
Here, $q_i$ and $\bar{q_i}$ stand for a generic quark and
the corresponding antiquark of the $i$-th flavour respectively.
$f_{\frac{q_{i}}{p}}$ ($f_{\frac{{\bar q_{i}}}{p}}$) is the    
quark (antiquark) distribution function within a proton.
$\tau \equiv {m_{g^{(1)}}^2 / S_{PP}}$, where $\sqrt{S_{PP}}$
is the proton-proton centre of momentum energy. 
$\Gamma(G^{1} \rightarrow q_i \bar q_i)$ represents the decay
width of $G^1$ into the quark-antiquark pair and is given by 
\begin{equation}
\Gamma = \left[\frac{{g^{2}_{G^{1}q q}}}{\pi}\right]m_{g^{(1)}} .
\end{equation}
Here $g^{}_{G^1 q q}$ is the KK-parity-violating coupling of the
$n = 1$ gluon with the SM quarks -- see Eqs. (\ref{coup1}) and
(\ref{coup2}). 

Eq. (\ref{x-sections}) represents the lowest order result in QCD.
We have not considered higher order contributions in our analysis
and used it bearing in mind that QCD corrections usually enhance
cross sections and so our results are probably conservative.

To obtain the
numerical values of the cross sections, we use a parton-level
Monte Carlo code with parton distribution functions as
parametrized in CTEQ6L \cite{CTEQ}.  We take the $pp$ centre of
momentum energy to be  8 TeV.  Renormalisation (for $\alpha_s$)
and factorisation scales (in the parton distributions) are both set at
$m_{g^{(1)}}$.

We are now ready to present the results of our investigation. To
our knowledge, this is the first of its kind where experimental
data from the LHC have been used to restrict the parameter space
of KK-parity-violating UED. Results for two distinct cases, either BLKTs are
present at both fixed points or only at one of the two, will be
presented in following two subsections.

\subsection{BLKT at both fixed points}

\begin{figure}[h]  
\begin{center} 
\includegraphics[scale=.36, angle=270]{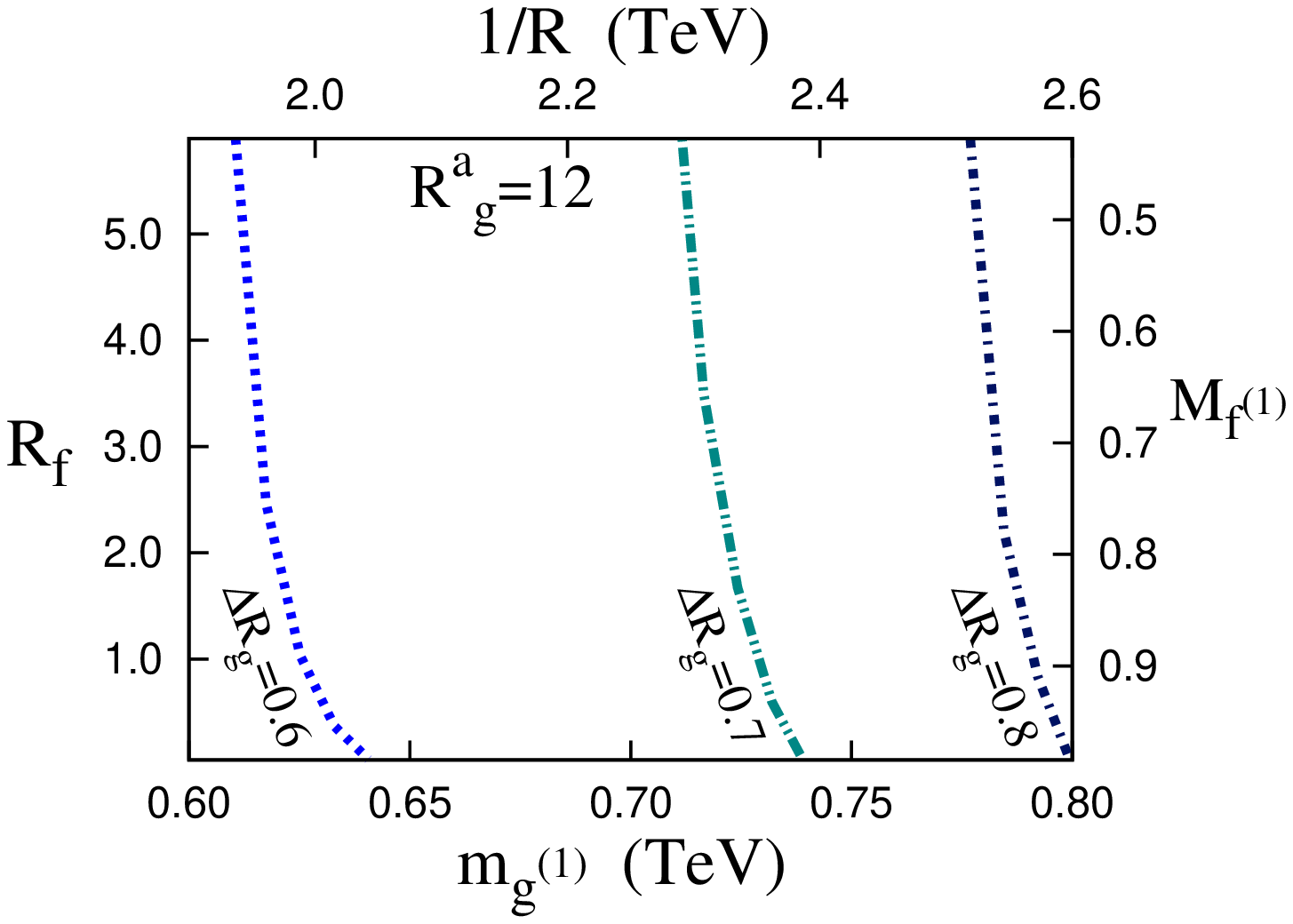}
\includegraphics[scale=.36, angle=270]{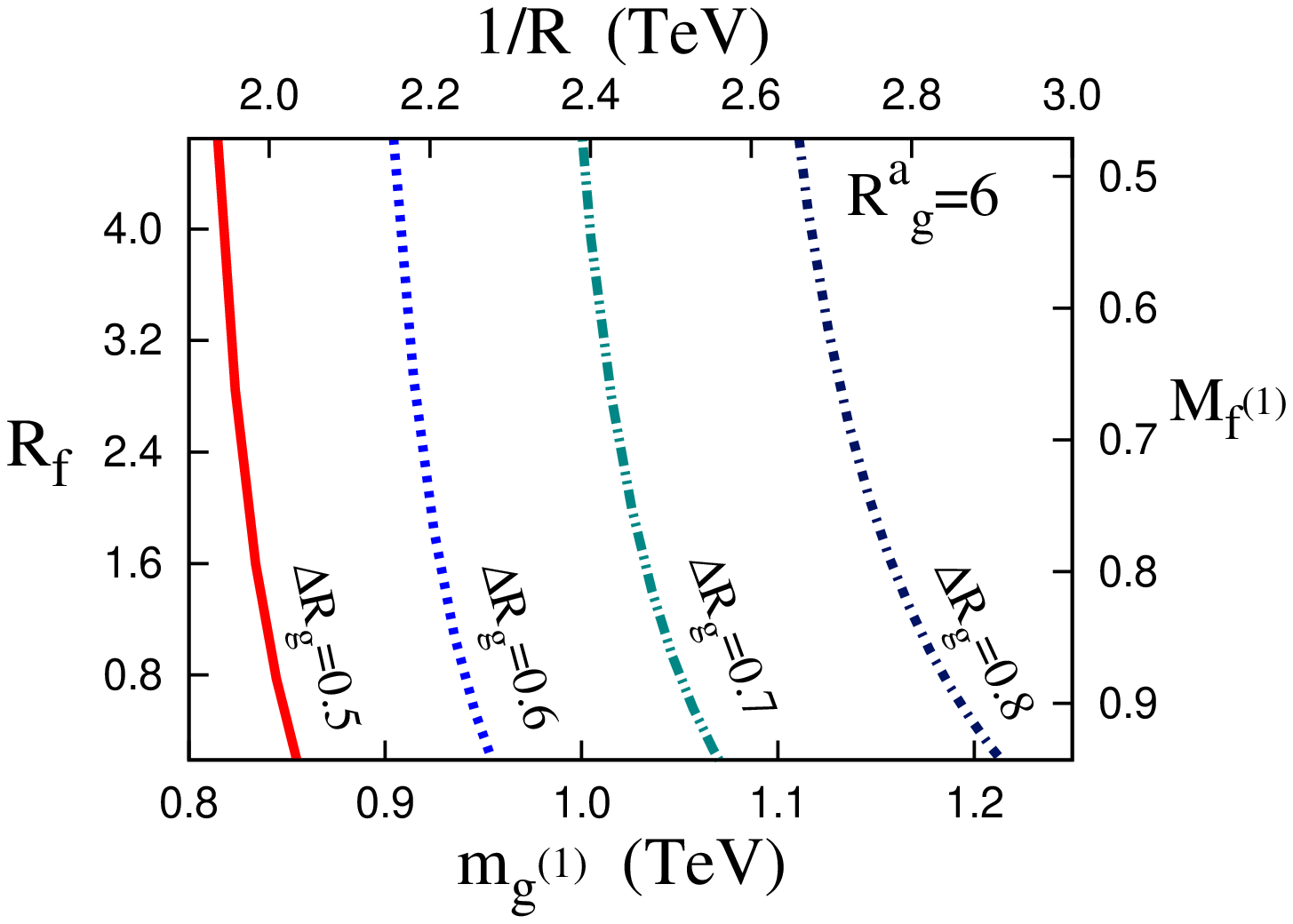}
\includegraphics[scale=.36, angle=270]{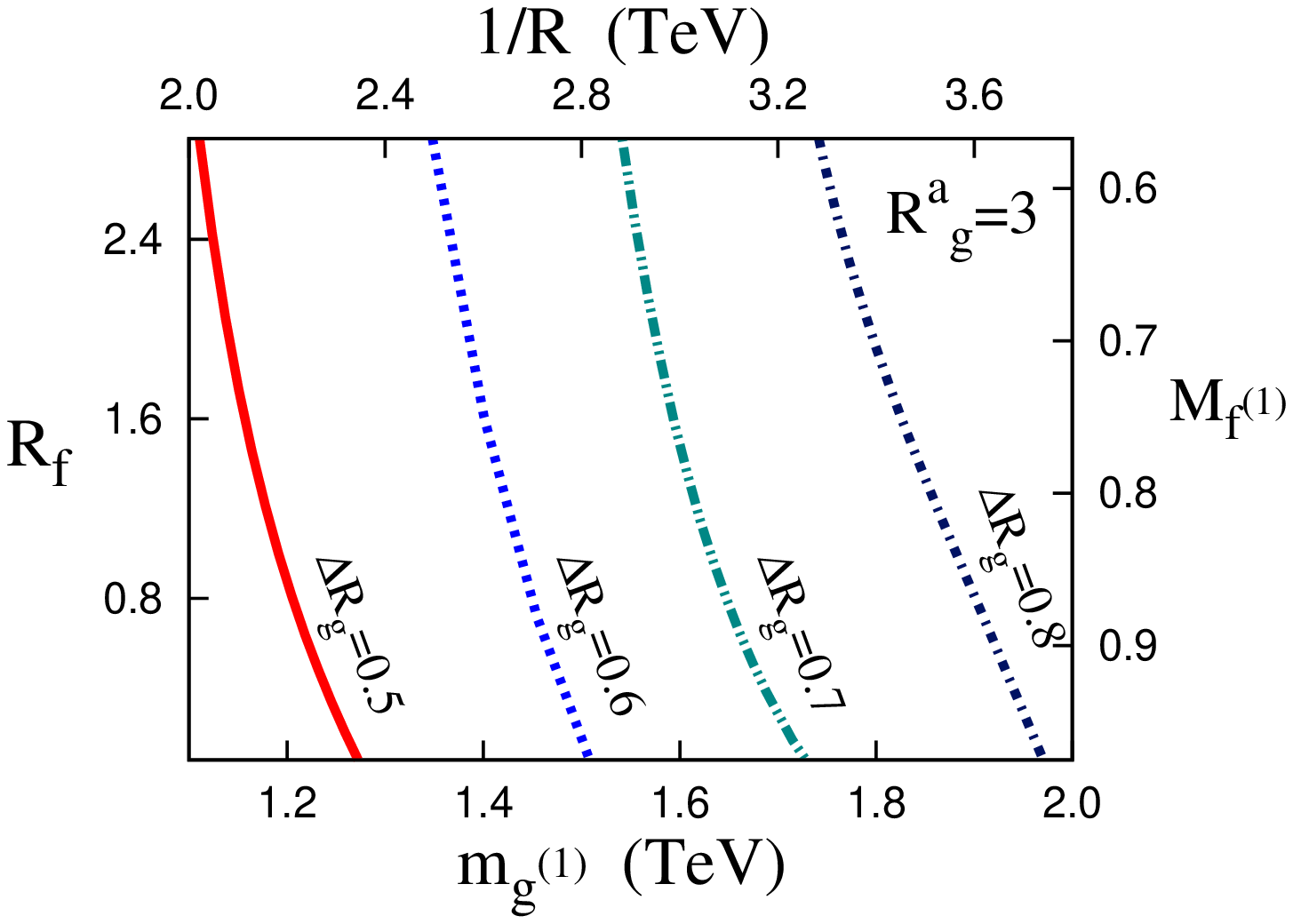}
\caption{95\% C.L. excluded/allowed regions in the $m_{g^{(1)}} - R_f$
parameter space for several choices of $\Delta R_g$ from
non-observation of a resonant $t\bar{t}$ signal at the LHC
running at 8 TeV. Each panel corresponds to a specific value of
$R^a_g$.  The region to the left of a given curve is disfavoured by the
CMS data \cite{cms8T}. $1/R$ and $M_{f^{(1)}} =
m_{f^{(1)}} R$ are displayed in the upper and right-side axes
respectively (see text).}
\label{contours} 
\end{center} 
\end{figure}

In Fig. \ref{contours} we show the region of parameter space
excluded by the  CMS 8 TeV data \cite{cms8T} for the case in
which the quark BLKTs are equal at the two fixed points but
KK-parity is broken by the unequal values of the gauge BLKTs.
The three panels correspond to different choices of $R_g^a$. In
each panel the region to the left of a curve  in the
$m_{g^{(1)}}-R_f$ plane is disfavoured by the  CMS  data.
The curves in any one panel correspond to different choices of
$\Delta R_g$ as indicated.

Since the KK-mass is rather insensitive to $\Delta R_g$,
for a chosen $R_g$ there is  one-to-one correspondence of
$m_{g^{(1)}}$ with $1/R$ which is shown on the upper axis of the
panels. Also, $R_f$ determines $M_{f^{(1)}} = m_{f^{(1)}} R$ and
is displayed on the right-side axis from where $m_{f^{(1)}}$
corresponding to any $1/R$ can be read off.
 
The results depicted  in Fig.  \ref{contours}
can be readily understood by noting that the LHC exclusion plots
translate in this model to a limit on the KK-parity violating
coupling for any chosen $n = 1$ gluon mass.  $R^a_g$ is fixed for
a panel. In any panel,  the 95\%
C.L.  CMS upper limit on the cross section times the branching
ratio  implies that the
points of intersection of the curves with a vertical line
corresponding to  a fixed $m_{g^{(1)}}$
identify ($R_f, ~\Delta R_g$) pairs which lead to the same
magnitude of the coupling constant.  From the left panel of Fig.
\ref{KKcoupling} it is seen that this happens if an increase in
$R_f$ is matched by an  increase in $\Delta R_g$. This is indeed
seen to be the case.  One can also easily explain the nature of
the plots from the standpoint of a fixed $R_f$. As one moves from
left to right keeping $R_f$ fixed,  i.e., to increasing 
$m_{g^{(1)}}$, the production cross section falls.
It can be checked that though the  CMS data also decreases
with increasing resonance mass this is slower than the kinematic
reduction.  Therefore in order to match the observed results a
larger KK-violating coupling is needed as $m_{g^{(1)}}$ increases. It is
seen from the left panel of Fig.
\ref{KKcoupling} that for a fixed $R_f$ the coupling grows as
$\Delta R_g$ is increased. This feature agrees with the results
in Fig. \ref{contours}.

The implications of the above results on the $n = 1$ level
KK mass spectrum can be extracted by considering them in
conjunction with Figs. \ref{KKmass} and \ref{KKcoupling}. The
limits on $m_{g^{(1)}}$ are essentially those on the $t\bar{t}$
resonance given in the data \cite{cms8T}. However, for any chosen
$R_g^a$ (any one panel) this entire range cannot be covered. This
is because the KK-parity violating coupling, which serves to
match the model prediction for the cross section for a particular
$m_{g^{(1)}}$ with the data, varies only over a limited range
(see left panel of Fig. \ref{KKcoupling}). This therefore
determines the KK-gluon mass band permissible for a certain
$R_g^a$. At the same time this puts an {\em upper} bound on the
$n = 1$ quark mass, which anyway has to be heavier than the $n =
1$ gluon in this model. Thus the quark KK  excitation mass
has to be in a limited range to agree with the LHC data.  This
feature can be 
illustrated by a few examples from Fig.
\ref{contours}. From the left panel ($R_g^a = 12.0$) one
finds that if $m_{g^{(1)}}$ = 625 GeV then $m_{q^{(1)}}$ is
bounded from above by 1.77 TeV. If, on the other hand
$m_{g^{(1)}}$ = 1.60  TeV the $n = 1$ quark is constrained (see
the right panel, $R_g^a = 3.0$) to lie between 1.60 and 2.32
TeV. In Fig. \ref{contours} the three $R_g^a$ choices cover
the entire CMS exclusion range of $t\bar{t}$ resonance mass.

\subsection{BLKT at one fixed point}

\begin{figure}[h] 
\begin{center} 
{\hskip -9cm}
\includegraphics[scale=0.50, angle=270]{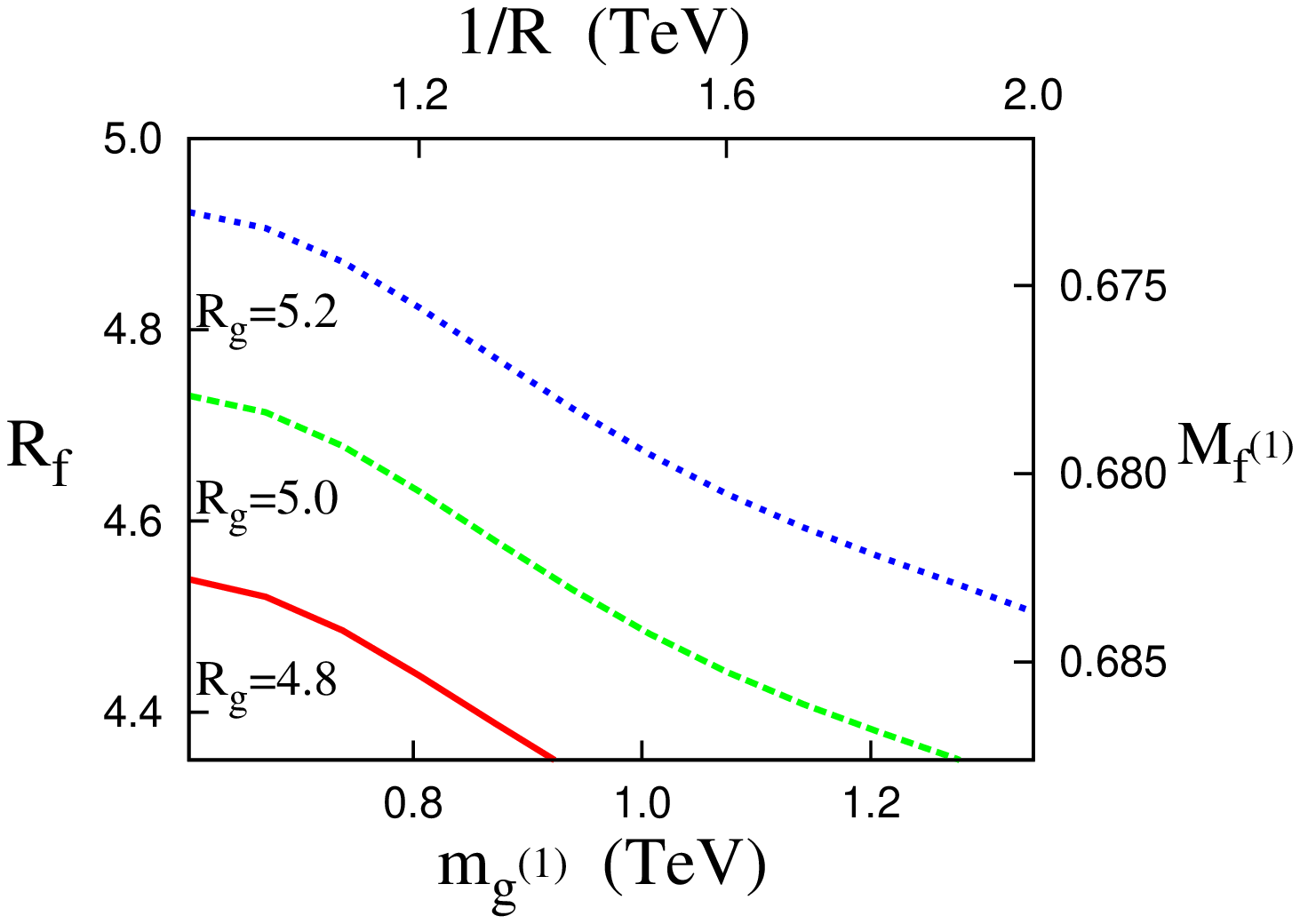}
{\vskip -5.4cm}
{\hskip 7cm}
\includegraphics[scale=.50, angle=270]{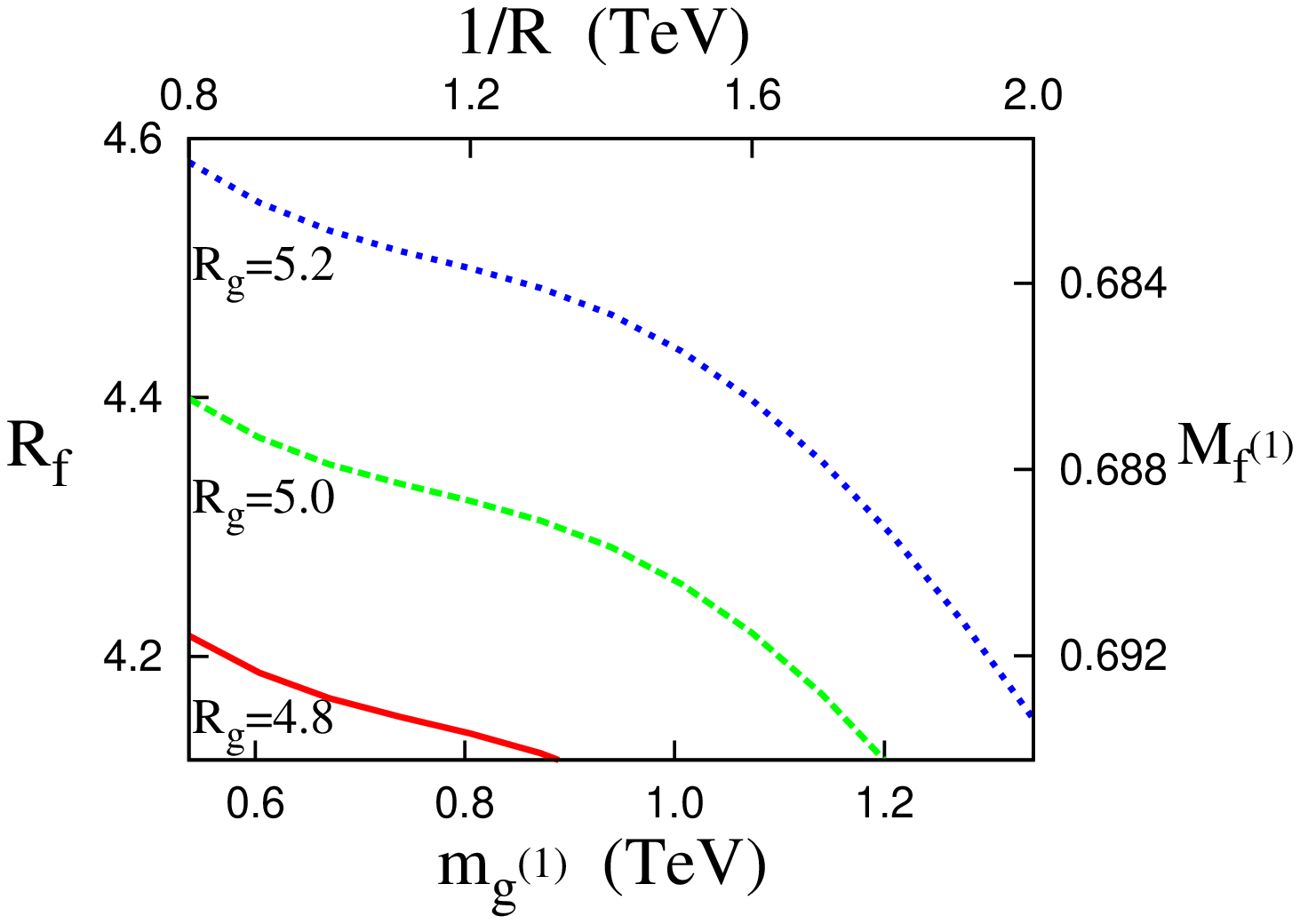}
\caption{95\% C.L. exclusion plots in the $m_{g^{(1)}} - R_f$ plane for
several choices of $R_g$. The region below a specific curve is
ruled out from the non-observation of a resonant $t \bar t$ signal
in the 8 TeV run of LHC by CMS \cite{cms8T} (left) and ATLAS
\cite{atlas8T} (right). $1/R$
and $M_{f^{(1)}} = m_{f^{(1)}} R$ are displayed in the upper and
right-side axes respectively (see text). The $y$-axis ranges
in the two panels are different.}
\label{contoursS} 
\end{center} 
\end{figure}

Now let us turn to the case of quark and gauge BLKTs at only
one fixed point.   In the left (right)
panel of  Fig. \ref{contoursS} we
show the bounds obtained using the 8 TeV results of CMS (ATLAS).
The relevant
KK-parity-violating couplings vanish when\footnote{Since for this
option the BLKTs are present at only one fixed point we denote
them by $R_f$ and $R_g$ with no superscript.} $R_f = R_g$. We
show in this case the exclusion  curves in the $m_{g^{(1)}}-R_f$ plane
for different choices of $R_g$. The region below a curve has
been ruled out from the data.  As expected the CMS bounds
based on 19.7 $fb^{-1}$ data are more restrictive than those from
the 14.3 $fb^{-1}$ ATLAS result.

Our discussion in the following is based on the left panel
of Fig. \ref{contoursS}. The nature of behaviour seen in Fig.
\ref{contoursS} tallies with our earlier considerations. As we
have noted in the  right panel of Fig. \ref{KKmass}, $M_{(1)}
\equiv m_{g^{(1)}} R$  is quite insensitive to the value of
$R_g$. It is a good approximation therefore to take the mass of
$G^1$ to be simply proportional to $1/R$; the $1/R$ values are
indicated in the upper axes of the panels in Fig.
\ref{contoursS}.  For any $m_{g^{(1)}}$ 
the CMS data gives a bound for the
corresponding cross section times branching ratio. Once the mass
is fixed, the experimental limit can be achieved by a specific
value for the KK-number violating coupling, i.e., the points of
intersection of a vertical line with the curves give $(R_g,
R_f)$ pairs which result in this fixed coupling. This can be
borne out by comparing with the right panel of Fig.
\ref{KKcoupling}. A second option is to consider the plots in
Fig. \ref{contoursS} for a fixed $R_f$. In this case as $m_{g^{(1)}}$
increases the enhanced mass hinders the production of the $n = 1$
KK gluon. We have checked that the fall seen in the  CMS data
with increasing resonance mass is slower then this kinematic
reduction. To compensate for this, the KK-violating coupling must
increase as we move to larger $m_{g^{(1)}}$. As seen from the right panel of
Fig. \ref{KKcoupling}, for a fixed $R_f$ the coupling enhancement
is accomplished by increasing $R_g$. This is the case in Fig.
\ref{contoursS}.

The conclusions that can be drawn from the above  results
are similar to the ones from Fig.
\ref{contours} but much more stringent.  For example, if
$m_{g^{(1)}}$ = 600 GeV then depending on whether $R_g$ is 4.8,
5.0, or 5.2 the upper bound on the $n = 1$ quark mass is 614, 610
or 605 GeV. For a heavier $n = 1$ gluon of mass 1.200 TeV one
finds that the upper bound on the corresponding quark excitation
is 1.236 for $R_g = 5.0$ or 1.227 TeV $R_g = 5.2$. These examples
indicate that in this scenario, the $n = 1$ quarks and gluons
have to be quasi-degenerate to tally with the LHC observations.

\section{Conclusions} 
Universal extra dimension models are among the
attractive options for beyond the standard model physics. 
Here the SM particles are complemented with heavier KK
excitations which are equispaced in mass. The interaction
strengths of these states are controlled entirely by the SM.
Various aspects of the model ranging from constraints from
precision measurements to collider searches have been
looked at in the literature. Signals for UED are being
actively searched for at the LHC.

One of the less attractive predictions of UED is the compressed
mass spectrum of KK excitations of all SM particles at any fixed
level. A remedy for this had been noted early on. It was shown
\cite{cms1} that five-dimensional radiative corrections split the
degeneracy in a calculable way determined by the SM charges of the
zero-mode states. The corrections are encoded as additional
four-dimensional interactions  located at the the two boundary
points (BLTs). In this version of UED, known as
minimal UED, the practice has been to assume that the couplings
of the KK excitations continue to be as for the SM particles and
only the mass degeneracy is removed.

In this work we examine departures of the boundary localized
kinetic terms from the above minimal choice. There are two
possibilities of choosing the four-dimensional kinetic terms at
the fixed points with rather distinct physics consequences. In
the first, the BLKTs are of equal strength at both fixed points
$(y = 0, \pi R$).  Here, a $Z_2$ symmetry $y
\longleftrightarrow (y - \pi R)$ survives.  One ends up with
a theory where the spectrum of KK-particles and the couplings can
be drastically different from mUED. The lightest among the $n =
1$ KK particles is stable and can be a dark matter candidate \cite{dm,ddrs_dm}.
The other alternative is to allow the BLKTs at the two fixed
points to be of unequal strengths.  This  will lead to a breakdown of
KK-parity and will allow, for example, single production of $n =
1$ KK-excitations and their decay to SM particles. Earlier  we have
examined, $B^1 (W_3^1) \rightarrow e^+e^-, \mu^+\mu^-$, decays
after the production of the $B^1 (W_3^1)$ singly at the LHC \cite{ddrs}.

In this article, we have considered the possible BLKTs for an
interacting theory of  quarks and gluons.
In one alternative, the strengths of quark BLKTs at the two
fixed points have been assumed to be equal $\equiv r_f$. For the
gauge boson boundary kinetic terms we have considered the general case of
unequal BLKTs $(r_g^a \neq r_g^b)$. Equality of the latter strengths
would restore a $Z_2$-parity. As an alternate possibility we have
considered the situation where the quark and gluon BLKTs are
present {\em only} at the $y = 0$ fixed point. In both cases the
boundary terms modify the field equations in the $y$-direction.
Consistency conditions of the solutions of the above equations
lead to the  masses of KK-excitations of quarks and the gluons
and their wave-functions in the $y$-direction.

We have calculated the coupling
of $G^1$, the $n = 1$ KK-excitation of the gluon, to a pair of
zero-mode quarks (i.e.,  SM quarks) as a function, in the first
alternative,  of $r_f, r^a_g,
r^b_g ~{\rm and} ~1/R$.  In general, we have presented the
coupling as a function of the scaled variable $R_f$  for several
choices of the other parameters.  A similar KK-parity-violating
coupling, which arises  when the BLKTs are present only at $y
=0$,  has also been evaluated.  The coupling is a hallmark of
KK-parity violation and vanishes in the $\Delta R_g = 0$ limit in
the first case and for $R_f = R_g$ in the second.

These results are utilized to calculate the production of $G^1$
singly at the LHC and its subsequent decay to $t \bar{t}$, both
production and decay being via the KK-parity-violating coupling.
The predictions are compared with the results on $t\bar{t}$
resonance production signature at the  LHC running at 8 TeV $pp$
centre of momentum energy  \cite{atlas8T, cms8T}.  It is revealed that
nonobservation of this signal with  19.7 $fb^{-1}$ accumulated
luminosity already disfavors a large part of the parameter space
(spanned by $r_f, r^a_g, r^b_g$ and $1/R$ in one case and $r_f,
r_g$ and $1/R$ in the other).  In the models considered here
the $n = 1$ gluon is lighter than the corresponding quark and the
bounds on the mass of the former are the same as that on the $t
\bar{t}$ resonance from the data. The cross section limits from
LHC put tight {\em upper} bounds on the $n = 1$ quark excitation mass. In
particular, while a range of a few hundred GeV  is still permitted
for this mass in the first scenario, in the second the $n = 1$
quarks and gluons have to be quasi-degenerate.

A similar new physics signal can also arise from other models,
e.g., if there are extra $Z$-like bosons as in the Left-Right
symmetric models or those with an extra $U(1)$ symmetry. Here we
have not attempted to compare the KK-parity-violating UED signals
with those from such other scenarios.

{\bf Acknowledgements} The authors are grateful to Prof. J.
Wudka, University of California, Riverside for asking a question
which led to this work. They thank Ujjal Kumar Dey for
collaboration in the early stages of this work. AD acknowledges
partial support from the DRS project sanctioned to the Department
of Physics, University of Calcutta by the University Grants
Commission. AR is thankful to the Department of Science and
Technology for a J.C.  Bose Fellowship. AS is the recipient of a
Senior Research Fellowship from the University Grants Commission.


\begin{thebibliography}{99} 


\bibitem{atlas7T}
The ATLAS Collaboration,  
 Phys.\ Rev.\ D {\bf 88} (2013) 012004 
[arXiv:1305.2756]. See also JHEP {\bf 09} (2012) 041.


\bibitem{atlas8T}
The ATLAS Collaboration,  
 [arXiv:1310.0486v2].


\bibitem{cms7T}
The CMS Collaboration, JHEP
{\bf 09} (2012) 029; JHEP
{\bf 12} (2012) 015;  Phys.\ Rev.\ D {\bf 87} (2013) 072002. 

\bibitem{cms8T}
The CMS Collaboration,  [arXiv:1309.2030v1]. 

\bibitem{acd} 
T.~Appelquist, H.~C.~Cheng and B.~A.~Dobrescu, 
 Phys.\ Rev.\ D {\bf 64} (2001) 035002 
 [arXiv:hep-ph/0012100]. 

\bibitem{georgi}
  H.~Georgi, A.~K.~Grant and G.~Hailu,
  Phys.\ Lett.\ B {\bf 506} (2001) 207
  [arXiv:hep-ph/0012379].

\bibitem{cms1}  
H.C.~Cheng, K.T.~Matchev and M.~Schmaltz,  
Phys.\ Rev.\ D {\bf 66} (2002) 036005  
[arXiv:hep-ph/0204342].  

\bibitem{cms2} 
 H.~C.~Cheng, K.~T.~Matchev and M.~Schmaltz, 
 Phys.\ Rev.\  D {\bf 66} (2002) 056006 
 [arXiv:hep-ph/0205314]. 




\bibitem{nath}  
P.~Nath and M.~Yamaguchi, 
Phys.\ Rev.\ D {\bf 60} (1999) 116006 
[arXiv:hep-ph/9903298].  
 
\bibitem{chk} 
D.~Chakraverty, K.~Huitu and A.~Kundu, 
Phys.\ Lett.\ B {\bf 558} (2003) 173 
[arXiv:hep-ph/0212047]. 
 
 
\bibitem{buras} A.J.~Buras, M.~Spranger and A.~Weiler,
Nucl.\ Phys.\ B {\bf 660} (2003) 225 
[arXiv:hep-ph/0212143]; 
A.J.~Buras, A.~Poschenrieder, M.~Spranger and A.~Weiler, 
Nucl.\ Phys.\ B {\bf 678} (2004) 455 
[arXiv:hep-ph/0306158]. 
 
\bibitem{desh}  
K.~Agashe, N.G.~Deshpande and G.H.~Wu, 
Phys.\ Lett.\ B {\bf 514} (2001) 309 
[arXiv:hep-ph/0105084];
  U.~Haisch and A.~Weiler,
  Phys.\ Rev.\ D {\bf 76} (2007) 034014
  [hep-ph/0703064].
 
\bibitem{santa}
  J.~F.~Oliver, J.~Papavassiliou and A.~Santamaria,
  Phys.\ Rev.\  D {\bf 67} (2003) 056002
  [arXiv:hep-ph/0212391].

\bibitem{appel-yee}
  T.~Appelquist and H.~U.~Yee,
  Phys.\ Rev.\ D {\bf 67} (2003) 055002
  [arXiv:hep-ph/0211023];
  G.~Belanger, A.~Belyaev, M.~Brown, M.~Kakizaki and A.~Pukhov,
  EPJ Web Conf.\  {\bf 28} (2012) 12070
  [arXiv:1201.5582 [hep-ph]].
 
\bibitem{ewued} T.G. Rizzo and J.D. Wells, Phys. Rev. D {\bf 61} 
(2000) 016007 [arXiv:hep-ph/9906234]; A. Strumia, Phys. Lett. B {\bf 
466} (1999) 107 [arXiv:hep-ph/9906266]; C.D. Carone, Phys. Rev. D {\bf 
61} (2000) 015008 [arXiv:hep-ph/9907362]. 

\bibitem{precision}
  I.~Gogoladze and C.~Macesanu,
Phys.\ Rev.\ D {\bf 74} (2006) 093012
[arXiv:hep-ph/0605207].

\bibitem{flacke2}L. Edelhauser, T. Flacke, and M. Kr\"{e}mer, 
JHEP {\bf 1308} (2013) 091
 [arXiv:1302.6076v2 [hep-ph]]. 
 

\bibitem{ddrs}
A. Datta, U. K. Dey, A. Shaw, and A. Raychaudhuri, Phys.
Rev. D {\bf 87} (2013) 076002.

\bibitem{Dvali}
  G.~R.~Dvali, G.~Gabadadze, M.~Kolanovic and F.~Nitti,
Phys.\ Rev.\ D {\bf 64} (2001) 084004
[arXiv:hep-ph/0102216].



\bibitem{carena}  
 M.~S.~Carena, T.~M.~P.~Tait and C.~E.~M.~Wagner, 
 Acta Phys.\ Polon.\ B {\bf 33} (2002) 2355 
 [arXiv:hep-ph/0207056]. 

\bibitem{delAguila}
  F.~del Aguila, M.~Perez-Victoria and J.~Santiago,
JHEP {\bf 0302} (2003) 051
[hep-th/0302023];  [hep-ph/0305119].

\bibitem{delAguila_STU}
  F.~del Aguila, M.~Perez-Victoria and J.~Santiago,
  Acta Phys.\ Polon.\ B {\bf 34} (2003) 5511
  [hep-ph/0310353].

\bibitem{flacke}  
 T.~Flacke, A.~Menon and D.~J.~Phalen, 
 Phys.\ Rev.\ D {\bf 79} (2009) 056009 
 [arXiv:0811.1598 [hep-ph]]. 




\bibitem{asesh}
  For a discussion of BLKT in extra-dimensional QCD, see
A.~Datta, K.~Nishiwaki and S.~Niyogi,
JHEP {\bf 1211} (2012) 154 [arXiv:1206.3987 [hep-ph]].

\bibitem{schwinn} 
 C.~Schwinn, 
 Phys.\ Rev.\ D {\bf 69} (2004) 116005 
 [arXiv:hep-ph/0402118]. 

\bibitem{CTEQ} J. Pumplin {\em et al.}, JHEP {\bf07} (2002) 012 
 [arXiv:hep-ph/0201195].

\bibitem{dm}  G. Servant and T. M. P. Tait, Nucl. Phys. B {\bf
650} (2003) 391, [arXiv: hep-ph/0206071];
D.~Majumdar, Mod.\ Phys.\ Lett.\ A {\bf 18} (2003) 1705;
K. Kong and K. T. Matchev, JHEP {\bf 0601} (2006) 038 [arXiv:
hep-ph/0509119];
F.~Burnell and G.~D.~Kribs,
Phys.\ Rev.\ D {\bf 73} (2006) 015001  [arXiv:hep-ph/0509118].


\bibitem{ddrs_dm}
A. Datta, U. K. Dey, A. Raychaudhuri, and A. Shaw, Phys.
Rev. D {\bf 88} (2013) 016011  [arXiv:1305.4507 [hep-ph]]. 

\end{thebibliography}
\end{document}